\documentclass[sigconf]{acmart}
\usepackage[british]{babel}
\usepackage[utf8]{inputenc}

\usepackage{placeins}
\usepackage{tikz}
\usepackage{booktabs}
\usepackage{amsmath}
\usepackage{breakurl}
\usepackage[nobiblatex]{xurl}
\usepackage{times,url,color,soul,xspace,enumitem}
\usepackage{graphicx}
\usepackage{multirow}
\usepackage{caption}
\usepackage{subcaption}
\usepackage{comment}
\usepackage{float}
\usepackage{xspace}
\usepackage{tikz}
\usepackage{amsmath}
\usepackage[inline,draft,nomargin,index]{fixme}

\usepackage{cleveref}
\crefformat{section}{\S#2#1#3} 
\crefformat{subsection}{\S#2#1#3}
\crefformat{subsubsection}{\S#2#1#3}

\FXRegisterAuthor{giu}{agiu}{\textcolor{red}{Giu}}

\FXRegisterAuthor{atr}{aatr}{\textcolor{blue}{Atr}}

\FXRegisterAuthor{lin}{alin}{\textcolor{green}{Lin}}

\begin{document}

\date{}

\title{\bf Users Really Do Respond To Smishing}

\author{Muhammad Lutfor Rahman}
\affiliation{%
  \institution{California State University San Marcos}
   \city{San Marcos}
   \state{CA}
   \country{USA}}
\email{mlrahman@csusm.edu}

\author{Daniel Timko}
\affiliation{%
  \institution{California State University San Marcos}
   \city{San Marcos}
   \state{CA}
   \country{USA}}
\email{timko002@csusm.edu}

\author{Hamid Wali}
\affiliation{%
  \institution{California State University San Marcos}
   \city{San Marcos}
   \state{CA}
   \country{USA}}
\email{wali003@csusm.edu}

\author{Ajaya Neupane}
\affiliation{%
  \institution{Palo Alto Networks}
   \city{Santa Clara}
   \state{CA}
   \country{USA}}
\email{aneupane@paloaltonetworks.com}

\renewcommand{\shortauthors}{Rahman, et al.}

\begin{abstract}
Text phish messages, referred to as \textit{Smishing} (SMS + phishing) is a type of social engineering attack where fake text messages are created, and used to lure users into responding to those messages. These messages aim to obtain user credentials, install malware on the phones, or launch smishing attacks. They ask users to reply to their message, click on a URL that redirects them to a phishing website, or call the provided number. Thousands of mobile users are affected by smishing attacks daily. Drawing inspiration by the works of Tu et al. (USENIX Security, 2019) on Robocalls and Tischer et al. (IEEE Symposium on Security and Privacy, 2016) on USB drives, this paper investigates why smishing works. Accordingly, we designed smishing experiments and sent phishing SMSes to 265 users to measure the efficacy of smishing attacks. We sent eight fake text messages to participants and recorded their CLICK, REPLY, and CALL responses along with their feedback in a post-test survey. Our results reveal that 16.92\% of our participants had potentially fallen for our smishing attack. To test repeat phishing, we subjected a set of randomly selected participants to a second round of smishing attacks with a different message than the one they received in the first round. As a result, we observed that 12.82\% potentially fell for the attack again. Using logistic regression, we observed that a combination of user REPLY and CLICK actions increased the odds that a user would respond to our smishing message when compared to CLICK. Additionally, we found a similar statistically significant increase when comparing Facebook and Walmart entity scenario to our IRS baseline. Following a similar approach, we identified that participants aged 18-24 had statistically lower odds of responding to our smishing scam when compared to participants aged 25-34. Alternatively, we found a statistically significant increase in the odds that African American and Mexican American participants potentially fall for our smishing scam against our baseline comparison. Lastly, participants who used their mobile devices for 11-20 hours a week had a statistically significant increase in the odds that they would potentially fall into smishing.
Based on our results, we pinpoint essentially message attributes and demographic features that contribute to a statistically significant change in the response rates to smishing attacks.
\end{abstract}

\maketitle
\section{Introduction}
Social engineering attacks are common forms of attacks within cybersecurity; they are rapidly increasing, and it cost billions of dollars each year~\cite{cyberfuture,whitehouse}. The mediums for social engineering attacks are emerging. Phishing, vishing, and smishing are different types of social engineering attacks that differ in how attackers contact the victims. More specifically, phishing uses email messages, vishing uses fraudulent phone calls, and smishing sends SMSes to the victim’s phone. There are classical studies~\cite{dhamija2006phishing,sheng2010falls,sheng2007anti,wu2006security} to understand why users are susceptible to email-based phishing attacks. However, there are a few or limited studies to understand users' susceptibility to vishing and smishing attacks, despite call-based or SMS-based email marketing having six to eight times of higher engagement rates than email-based marketing~\cite{eztextingstats} and smartphone users being three times more likely to get into a web-based phishing attack than desktop users~\cite{alkhalil2021phishing}.

Attackers are increasingly targeting billions of users~\cite{statista} on their smartphones, unlike desktops and laptops, to obtain sensitive information or infiltrate their associated organizations. Furthermore, they primarily attack mobile devices by targeting installed applications used for daily activities. However, organizations’ increasing use of automated text messages for notifications such as appointment confirmations or delivery notifications have presented opportunities for malicious actors to craft realistic phish messages. Smishing may lead to ransomware, spyware, or adware installation~\cite{bitdefender} on the user’s mobile device, which could allow an attacker to access personal data, e.g., passwords, social security numbers, contact information, banking credentials, and location data~\cite{scandolabodily,nagunwa2014behind}. 

Tu et al.~\cite{tu2019users} conducted a systematic Robocall study with university employees and faculties to systematically explore fraudulent phone calls. In the Robocall study, the authors developed an automatic service to call telephone numbers and retrieve users’ SSN. Smishing attacks differ from Robocall scams. In Robocall, attackers call the victim’s numbers and attempt to retrieve personal information. On the other hand, in smishing, attackers send messages to users containing vital information or fraudulent URLs—or both together. As per the RoboKiller Smart Phone app~\cite{robocall}, attackers sent 87 billion spam texts (compared to 72 billion spam calls) in 2021. Based on the Robokiller report, several factors are involved in this shift. First, due to COVID-19, call center-based hacking operations shifted to home. As a result, hackers primarily shifted to messaging. Second, the communication preference of Millennials plays a vital role in this shift. As per the Openmarket survey, 75\%~\cite{Openmarket} of Millennials prefer texts to calls and 83\%~\cite{eztextingstats} of them open SMS messages within 90 seconds of receiving them. Third, the effect of Anti-Robocall standards (i.e., STIR/SHAKEN framework). The Federal Communications Commission (FCC) mandated that STIR/SHAKEN framework went into effect on June 30, 2021~\cite{fcc}. Accordingly, it is expected that the gap between the number of spam text messages and Robocalls will also widen within the upcoming years.

Smishing attacks are growing at an alarming rate. Within the first six months of 2021, smishing increased by 700\%~\cite{smishincrease}. Even though smishing is a growing problem, very little research has been done to study who falls for these attacks, why they fall for them, and how carriers behave under smishing. The most relevant smishing study has been carried out by Blancaflor et al.~\cite{blancaflor2021let}. Within their study, they created a phishing website and sent it to 24 targeted users via SMS. They found that 1 out of 24 targeted users clicked the phishing URL. Unlike the study by Blacaflor et al., we prepared eight different types of messages. We sent them to 265 unique users, modifying different communication attributes to analyze what influences whether the smishing is successful (Ref~\ref{subsec: attributes}). 

In this paper, we set out to enhance the current knowledge of mobile users' susceptibility to smishing. Our study builds upon the existing framework~\cite{blancaflor2021let,tu2019users} to investigate the effects of different smishing attributes and demographics of successful smishing attacks. Smishing scams can utilize different techniques and methods to provide unique identification challenges to users that set them apart from Robocalls, which scammers use to trick their targets to fall for their smishing attacks~\cite{leonovsocialengineeringtechniques,yeboah2014phishing}. Intrinsic to the messaging system, SMS messages can provide additional obfuscation, for example, spoofed phone numbers and organization names for the attacker to trick their target into obtaining the victim’s response as well as access to fraudulent websites and shortened links. In this study, we introduced attributes, such as user response actions, message contents, entity, and area code, to identify their impact on the success rate of smishing scams in a diverse demographic of participants. To explore smishing attacks empirically and systematically, we formulated the following research questions:

\begin{itemize}
    \item \textbf{RQ1} What is the success rate of smishing? 
    \item \textbf{RQ2} Do different attributes of messages have an impact on the success rate of smishing? 
    \item \textbf{RQ3} How does smishing impact different demographics?

\end{itemize}

To answer these questions, we studied which user demographic affects the success rate of smishing attacks. We designed our experiments with different contexts, and smishing attributes to understand what would persuade a user to disclose their personal information. With the results from these experiments, we can infer how often carriers block smishing messages, how often users fall for them, and what solutions we require to combat this problem.

\vspace{3mm}
\textbf{Our Contributions:} In this paper, we present the results of a real smishing attack study with consent from the participants who were students from our university as well as the general public from different social media platforms. Our contributions are summarized below:

\begin{itemize}

    \item This is a comprehensive empirical study on the impact of different attributes on smishing and demographics in a near realistic scenario. We designed, developed and conducted a user experiment where users were subjected to smishing attacks without their prior knowledge of smishing study. In this paper, we summarize our experiments with 265 unique participants on twelve different smishing texts with greed and fear as motivational factors. In the first round we delivered smishing messages to 260 participants, and 16.92\% of them responded to the text messages (i.e., either called, clicked, or replied) ( Ref Table~\ref{table:summarresults}). We delivered smishing messages to 195 participants in the second round of smishing attempts. At that stage, 12.82\% of users fell for the second smishing attempt (Ref Table~\ref{table:summarresults}). We also perform a logistic regression to determine the impact that each demographic feature has on the odds that a participant potentially falls into smishing scams. The results, which can be found in Table~\ref{tab:demographicsLogistic}, show statistically significant impacts in the factors of age, ethnicity, and mobile usage when compared to our majority case baseline. 
    
    \item We designed text SMS experiments using stimuli with entity, scenario, user action, and motivation variations which can be found in Table~\ref{table:summarresults}. We compared these experiments using a Mann-Whitney U test across each type of attribute. These tests revealed that entity changes in our fear-based scenario had a statistically significant effect on potential smishing rates (Ref Section~\ref{tab:design}). Additionally, we performed a logistic regression on message attributes in Table~\ref{tab:AttributeComparison}. Our results show that when compared to our majority cases, user action and entity had a statistically significant impact on smishing rates.
    
    \item In the post-test survey, we asked our participants what the convincing factors for responding to the text SMS were. We found that most of them thought the SMS were real as they seemed to come from a government agency or a local area code (Ref Section~\ref{sec:survey}).
    
\end{itemize}

\textbf{Study Implications} We need to understand smishing as there has been a rise in the usage of text notifications for information dissemination from organizations and, thus, phishing via text messages from attackers. Previous research has shown that smartphone users are more likely to get into phishing attacks than desktop users. This study can provide an empirical foundation for measuring the effects of smishing attributes and demographic information on smishing success. Based on the variance in smishing rates across participant demographics, we believe that a deeper exploration into how demographics, combined with changes in smishing attributes, could lead to better explanations for why users fall into smishing scams. Additionally, in terms of future research, we found that our smishing messages, which were crafted to mimic realistic smishing messages, had a high rate of smishing success. However, when it came to collecting survey feedback, few participants were willing to respond to what caused them to fall for the message. Therefore, exploring why participants are unwilling to talk about their experience, even when debriefed, could provide important insights into how to effectively educate the public about smishing attacks. Based on the smishing rate, it is expected that carriers or third parties should develop robust mechanisms for smishing detection. The users should be aware of smishing attacks and their subsequent consequences. Our study will help us understand to design users' training materials, notification messages, and defensive mechanisms to augment users’ decision-making against such attacks. 

\section{Study Design \& Data Collection}

\begin{figure*}[htbp]
    \centering
        \includegraphics[clip, trim=4cm .1cm 2cm .1cm, width=.68\textwidth]{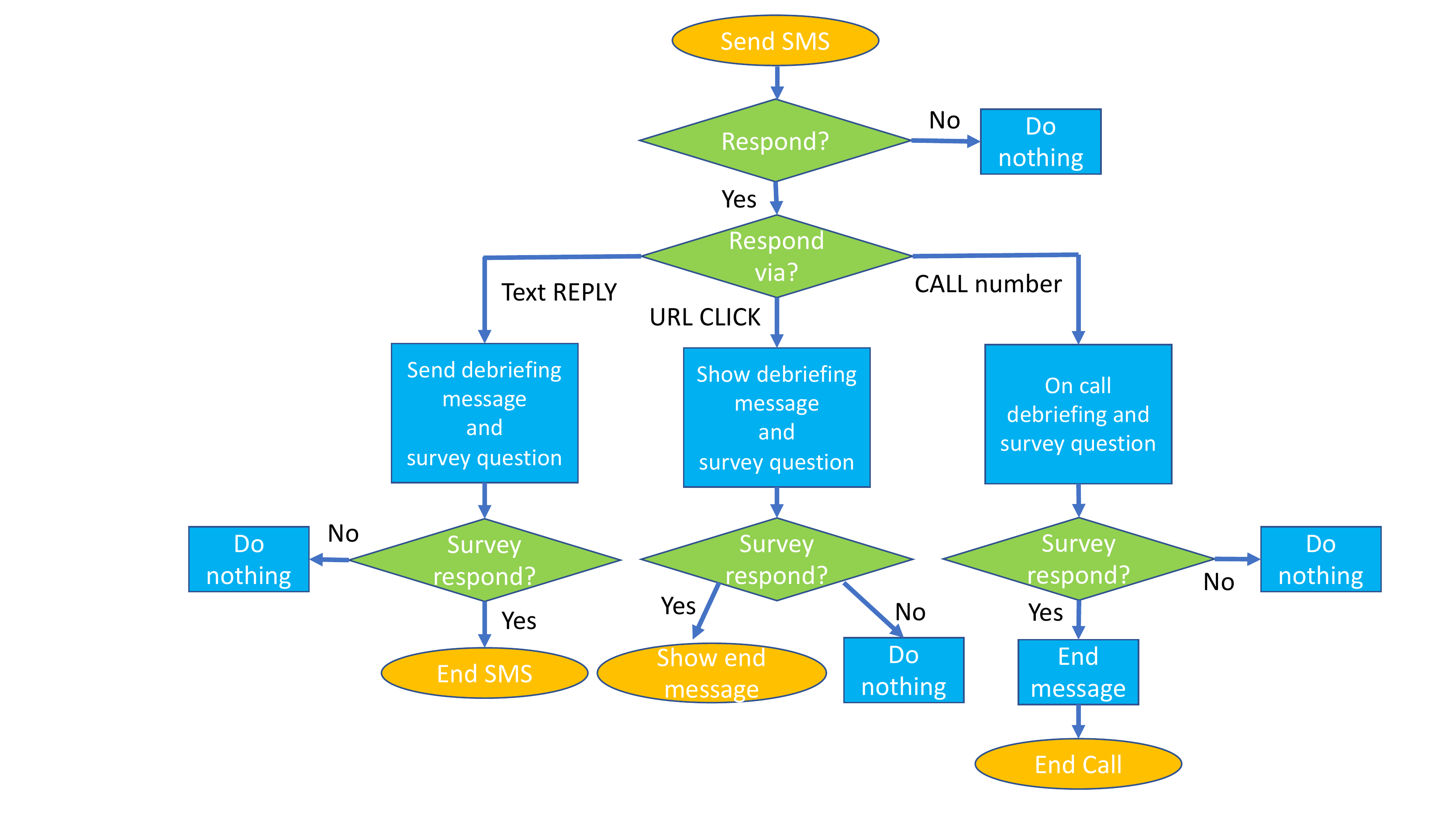}
    \caption{A flow chart of the experiment procedure.}
    \label{fig:catprocessflow}
\end{figure*}

Our research study is inspired by two prior studies on Robocalls~\cite{tu2019users} and USB drives~\cite{tischer2016users}. Additionally, our study aims to determine why users respond to smishing attempts. To execute our study, we first selected a set of attributes that we could manipulate in our smishing messages. Next, we designed twelve experiments consisting of different variations of these attributes. Then, we organized these experiments based on the attribute types, and motivation to run statistical analysis. The experiments followed a strict flow chart procedure, dictating how to respond to user actions. This flow chart can be seen in Figure~\ref{fig:catprocessflow}. Furthermore, we set our experiments to study the impact demographics and attributes may have on smishing success rates. The understanding of demographics, including race and ethnicity, helps us understand the unfairness and bias that may exist in the preparation of security materials. A full list of the attributes and demographics along with their types and descriptions can be found in Table~\ref{tab:DemographicsAttributesDefined}. While we were deceptive, this study was performed by the approval of our university’s IRB and went through great lengths to accommodate ethical concerns.

\subsection{Attributes}
\label{subsec: attributes}

To choose our attribute set, we carefully reviewed different attributes used in previous literature~\cite{tu2019users}. In addition, we also explored possible attributes that could be chosen from smishing messages, and information posted on different Internet sources, news, websites, YouTube, and user comments. We describe the attributes chosen for this research in the following sections.

\vspace{3mm}
\textbf{Entity}: One of the attributes we choose was entity. We define entity as an organization or subject intending to reach a user with a message. For example, one of the entities we used is the scammer pretending to be associated with a “Bank.” In contrast, in another experiment, the attacker pretended to be from “Facebook’s Security Team,” informing a user about a login attempt on their Facebook account from a different location and providing a fake link to change their password. By using different entities, we can compare to see what convinced the victim to reply. We also confirmed this by asking the victim to respond to survey questions at the end of the experiment, once they are debriefed, to discover what convinced them.

\vspace{3mm}
\textbf{Scenario}: Another attribute we used was the scenario. We define scenario as the content of the message sent to the victim user. An example of a scenario could be a message claiming that the victim’s social security number has been suspended or a message requesting approval for a bank transaction. We focused on two different scenarios for this study: reward and fear. A reward is something a victim is receiving in recognition of their effort or achievement, and fear is a stimulus that triggers the perception of danger or punishment. By studying varying scenarios, we can understand what convinces the victim to respond to the smishing message—are they more prone to a reward or a fear scenario?

\vspace{3mm}
\textbf{Area Code}: The area code from which a user receives a text may also influence the victim’s decision to either reply to a message or not. Scammers consciously select an area code with a scenario and entity they are using to scam a victim. For example, IRS scammers use the (202) area code in their caller IDs (i.e., associated with Washington, DC) to make it appear as if the IRS is calling or texting. In our experiments, we use multiple area codes, including local area codes near the victim’s location and toll-free codes like (833) or (855).

\vspace{3mm}
\textbf{Method}: The typical approach for smishing is to send an SMS to the victim’s phone number. There are three ways a victim can fall for smishing; these include:  

\begin{itemize}
    
    \item \textit{Click the URL:} This is the most convenient way of smishing. In this method, the victim is prompted to click a link that redirects them to a fake website which asks them to submit information into a phishing form—this is controlled by the malicious actor. A hacker can install malware or spyware using this method as well. Throughout this study, we refer to this method as CLICK.

    \item \textit{Reply through text:} Another convenient way of smishing is to reply through text. In this method, hackers collect sensitive information via text reply. Throughout this study, we refer to this method as REPLY.

    \item \textit{Call back on the provided number:} The last method of smishing is to call a provided number that provided along with the spam messages. In this way, the hackers collect personal and financial information from the victim. Throughout this study, we refer to this method as CALL. 
\end{itemize}

\begin{table*}[]
\centering
\small
\begin{tabular}{@{}llll@{}}
\toprule
\textbf{No.} & \textbf{Description} & \textbf{Type} & Baseline \\ \midrule
 & Message Attributes &  &  \\ \midrule
1 & Area Code: The 3 digits which refers to the service region that smishing messages were sent from. & Categorical & Local \\
2 & Entity Scenario: The organization or subject that we are attempting to impersonate in our message. & Categorical & IRS \\
3 & Scenario: The content of the message which falls into categories of smishing schemes. & Categorical & Lawsuit \\
4 & User Action: The action by which the smishing attack is attempting to persuade the victim to perform. & Categorical & Click \\
5 & Motivation: We use fear or rewards to try to incentivise our targets to take action. & Boolean & Fear \\ \midrule
 & Demographics &  &  \\ \midrule
6 & Gender: The gender of the participant. & Boolean & Male \\
7 & Age: The age of participants measured in categorical ranges. & Categorical & 24-35 \\
8 & Ethnicitiy: The ethnicity of participants in the study. & Categorical & White \\
9 & Weekly Mobile Usage: The number of hours the participants use their phone per week. & Categorical & 6-10 hours \\
10 & Education: The highest level of educational degree attained by the participant & Categorical & Bachelors degree. \\
11 & Profession: The current job of the participant. & Categorical & Student \\ \bottomrule
\end{tabular}
\caption{The types and descriptions for attribtues and demographics in the study. Baseline selected based on the majority case. }
\label{tab:DemographicsAttributesDefined}
\end{table*}

\begin{table*}[]
\centering
\small
\begin{tabular}{llccclll}
\hline
\textbf{No.} & \textbf{Message} & \textbf{SMS sender ID} & \textbf{Area code location} & \multicolumn{1}{r}{\textbf{Entity Scenario}} & \textbf{Scenario} & \textbf{User Action} & \textbf{Motivation} \\ \hline
E1 & M1 & 1-833-222-9721 & Toll-free & Walmart & Gift Card & CLICK & Reward \\
E2 & M2 & 1-914-529-6528 & Westchester, NY & Bank & Bank Transaction & REPLY & Fear \\
E3 & M3 & 1-xxx-xxx-1803 & Local & Law Enforcement & Arrest Warrant & CALL & Fear \\
E4 & M4 & 1-855-777-8616 & Toll-free & Facebook & Account Login & CLICK + REPLY & Fear \\
E5 & M5 & 1-xxx-xxx-1803 & Local & IRS & Lawsuit & CALL & Fear \\
E6 & M6 & 1-xxx-xxx-1803 & Local & Video website & Private video & CLICK & Fear \\
E7 & M7 & 1-833-222-9721 & Toll-free & FORTUNE 500 Co. & Job offer & REPLY & Reward \\ \hline
E8 & M1 & 1-833-222-9721 & Toll-free & Walmart & Gift Card & CLICK & Reward \\
E9 & M3 & 1-855-777-8616 & Toll-free & Law Enforcement & Arrest Warrant & CALL & Fear \\
E10 & M5 & 1-855-777-8616 & Toll-free & IRS & Lawsuit & CALL & Fear \\
E11 & M7 & 1-914-529-6528 & Westchester, NY & FORTUNE 500 Co. & Job offer & REPLY & Reward \\
E12 & M8 & 1-xxx-xxx-1803 & Local & FedEx & iPhone 13 & CLICK & Reward \\ \hline
\end{tabular}
\caption{Summary of all experiments. We have conducted E1-E7 in the first round and E8-E12 in the second round.}
\label{table:numbersummary}
\end{table*}

\subsection{Experimental Design}
\label{tab:design}

We designed our experiments to study the impact of different attributes (as defined in section~\ref{subsec: attributes}) on the success of smishing attacks. As illustrated in Table~\ref{table:numbersummary}, there are twelve experiments in total. For each experiment, we sent out one of the 8 unique text messages to participants. Each text message contained a combination of entity, scenario, area code, user action, and motivation. We devised multiple attributes to study how they might affect a user’s willingness to respond. For example, a person might be more likely to reply to a bank transaction scenario compared to a scenario where a gift card is being offered.

\begin{table*}[h]
\centering
\footnotesize
\begin{tabular}{@{}ll@{}}
\toprule
\multicolumn{1}{c}{\textbf{No.}} & \multicolumn{1}{c}{\textbf{MESSAGE CONTENT}} \\ \midrule
\textbf{M1} & \begin{tabular}[c]{@{}l@{}}Congratulations! You’ve just won a \$500 Walmart gift card due to your last purchase at Walmart this Thanksgiving. Claim your gift card here:\\ https://bit.ly/walmart-claimgiftcard\end{tabular} \\ \midrule
\textbf{M2} & \begin{tabular}[c]{@{}l@{}}ALERT from your Bank Account: Did you just spend \$289.15 at Walmart. If you attempted this transaction, reply YES.  If you do not recognize\\ this transaction, reply NO\end{tabular} \\ \midrule
\textbf{M3} & \begin{tabular}[c]{@{}l@{}}Due to some suspicious activities on your name, the Local Law Enforcement Agency is going to file an arrest warrant against you. \\ To get more information, call us as soon as you can on this number (XXX) XXX-1803\end{tabular} \\ \midrule
\textbf{M4} & \begin{tabular}[c]{@{}l@{}}The system detects that someone from an IP 2600:1800:b190:6850:5b5:b28e:9a61:baa5 ( America/Los Angeles) logged in to your Facebook\\ account that has never been used before. If it is you, reply immediately with YES. If it is not you, reply NO. To opt-out of text security alerts,\\ reply STOP. Please log in to your account to change the password immediately. https://fecebook.co/login Facebook Security Team\end{tabular} \\ \midrule
\textbf{M5} & \begin{tabular}[c]{@{}l@{}}We tried contacting you through mail several times and have had no responses. The IRS investigation team found some financial activities on\\ your name which were not matched on your last year’s tax return information. The IRS is going to file a lawsuit against you due to this discrepancy. \\ To get more information, call us as soon as you can on this number (XXX) XXX-1803\end{tabular} \\ \midrule
\textbf{M6} & Is it you in the video? https://www.xyxtube.com/watch?v=nt4BGsC9Vc8 \\ \midrule
\textbf{M7} & \begin{tabular}[c]{@{}l@{}}Hello, My name is Harris and I represent a Fortune 500 company. I would like to interest you in a chance to earn \$5000 a month in the comfort of your\\ own home. If you are interested, reply YES and we can discuss more of this opportunity. If you decline this opportunity, reply NO. To opt-out of text alerts,\\ reply STOP.\end{tabular} \\ \midrule
\textbf{M8} & FedEx: Your number won a brand new Apple iPhone 13, Pro Max. Please claim ownership and confirm the delivery address here. https://bit.ly/fedextrk \\ \bottomrule
\end{tabular}
\caption{List of smishing SMS for our experiment.}
\label{table:messagingcontent}
\end{table*}

In Table~\ref{table:messagingcontent}, you can see the comprehensive list of messages designed for this experiment using the three user-action methods that were previously defined. Each message incorporates one or more user actions to handle the response from users and gauge whether they are victims of smishing. All messages are designed to be sent through SMS texting with a pre-defined user-action response tied to each message. However, we record any REPLY, CLICK, or CALL responses to our smishing. 

To handle the user actions for our smishing messages, we used a flowchart procedure with each participant; this can be found in Figure~\ref{fig:catprocessflow}. As seen in the flowchart, we utilize a similar debriefing procedure to the study by Tu et al.~\cite{tu2019users}. Participants who respond to the text messages are debriefed and we ask them survey questions. Due to each user-action method being handled over a different medium, we also incorporate three different approaches to record their survey responses. Therefore, if we ask the victim to reply through text and they do reply, our next course of action would be to debrief them and send the survey question. Once the user responds to our survey question, we finally send the ending message. If we ask the victim to click on a URL, they will be redirected to a Qualtrics website where they can immediately view the debriefing message and survey question where a dialogue box will be provided for them to answer the question. Finally, the ending message will be displayed. If we ask the victim to call back on a number provided in the text message, once the user calls the number, a pre-recorded debriefing message and survey question will be played. Once the user answers our survey question, the ending message will be played, and the victims survey answer will be recorded as part of  data collection. In our research study, we do not ask for or collect any personal information or data during our experiment. Our goal is to see whether victims fall for the scam. Consequently, by calling back, clicking on a link, or replying to our text, they will let us know if they have fallen for the scam. Therefore, there is no reason for us to collect personal information from the participants and the IRB guidelines had to be followed. Another reason we are conducting this experiment is to find out how different attributes can convince victims to fall for the scam and later make comparisons between these attributes. This will provide insight into what convinces a human mind. For example, there might be a notion of fear in our text message, which might convince them to reply or call back.

\section{Study Procedures}
In this section, we provide details on how we mitigate the ethical issues of our experiments. We also describe participants' recruitment for our study.

\subsection{Ethical and Safety Consideration}
Participation in our experiment was voluntary. We advertised our study by stating that we were conducting a behavioral study for mobile users to measure the efficacy of information conveyed via text messages. However, we did not mention that we would send smishing SMS to their phone numbers. In addition, we did not provide the sender’s number. Due to the deception in our study, we carefully designed our experiments. We worked with our university’s Institutional Review Board (IRB) to get their approval and minimize harm. To maximize realism of the experiment, a good deal of prominent privacy and security related user studies~\cite{salah2012deceive,schecter2007emperor,adar2013benevolent, egelman2010tell} have used or advocated the use of deception. The use of deception may provoke an ethical debate, but it can be reasonably safe if we carefully design the ethical aspects of the experiment. To protect our participants, we inserted multiple safeguards in our experimental design.

\begin{table*}[h]
\centering
\small
\begin{tabular}{@{}lcccccccccc@{}}
\toprule
\textbf{No.} & \textbf{Message} & \textbf{\#Attempts} & \multicolumn{1}{l}{\textbf{\#Sent}} & \textbf{\#Undelivered} & \textbf{\#Delivered} & \textbf{CALL} & \textbf{REPLY} & \textbf{CLICK} & \textbf{Total} & \textbf{\%Success Rate} \\ \midrule
E1 & M1 & 86 & 4 & 52 & 30 & 1* & - & 8 & 9 & 30.00\% \\
E2 & M2 & 84 & 0 & 34 & 50 & - & 6 & - & 6 & 12.00\% \\
E3 & M3 & 75 & 0 & 44 & 31 & 1 & 2* & - & 3 & 9.68\% \\
E4 & M4 & 85 & 4 & 55 & 26 & 1* & 0 & 8 & 9 & 34.62\% \\
E5 & M5 & 96 & 0 & 32 & 64 & 5 & 3* & - & 8 & 12.50\% \\
E6 & M6 & 105 & 0 & 75 & 30 & - & - & 4 & 4 & 13.33\% \\
E7 & M7 & 91 & 4 & 58 & 29 & 1* & 4 & - & 5 & 17.24\% \\ \midrule
\textbf{1st Round Total} &  & 622 & 12 & 350 & 260 & 9 & 15 & 20 & 44 & 16.92\% \\ \midrule
E8 & M1 & 50 & 7 & 2 & 41 & 1* & - & 4 & 5 & 12.20\% \\
E9 & M3 & 50 & 1 & 40 & 9 & 0 & - & - & 0 & 0.00\% \\
E10 & M5 & 50 & 10 & 3 & 37 & 2 & - & - & 2 & 5.41\% \\
E11 & M7 & 49 & 0 & 2 & 47 & - & 7 & - & 7 & 14.89\% \\
E12 & M8 & 99 & 0 & 39 & 60 & 1* & - & 10 & 11 & 18.33\% \\ \midrule
\textbf{2nd Round Total} & \multicolumn{1}{l}{} & 298 & 18 & 86 & 195 & 4 & 7 & 14 & 25 & 12.82\% \\ \midrule \midrule
\textbf{Grand Total} & \multicolumn{1}{l}{} & 920 & 30 & 436 & 454 & 13 & 22 & 34 & 69 & 15.20\% \\ \bottomrule
\end{tabular}
\caption{The final delivery status for unique messages in the experiment. Here, * means the message did not have this user action,
but received responses from participants. - means user actions were not mentioned during that experiment. Note we only consider \#Delivered in our potential smishing rates.}
\label{table:summarresults}
\end{table*}

\begin{table*}[h]
\centering
\small
\begin{tabular}{@{}lccccccccccccc@{}}
\toprule
 & \multicolumn{12}{c}{Experiments} & \multicolumn{1}{l}{} \\ \midrule
 & E1 & E2 & E3 & E4 & E5 & E6 & E7 & E8 & E9 & E10 & E11 & E12 & \textbf{Total} \\ \midrule
Error   Code:30003 & 3 & 9 & 16 & 9 & 4 & 7 & 7 & 2 & 0 & 1 & 0 & 0 & 58 \\
Error Code:30005 & 2 & 0 & 0 & 6 & 0 & 1 & 8 & 0 & 0 & 2 & 0 & 0 & 19 \\
Error Code:30006 & 47 & 25 & 28 & 40 & 28 & 67 & 43 & 0 & 0 & 0 & 2 & 5 & 285 \\
Error Code:30007 & 0 & 0 & 0 & 0 & 0 & 0 & 0 & 0 & 40 & 0 & 0 & 34 & 74 \\ \midrule
\textbf{Error Code Sum} & 52 & 34 & 44 & 55 & 32 & 75 & 58 & 2 & 40 & 3 & 2 & 39 & 436 \\ \bottomrule
\end{tabular}
\caption{The status of all unique undelivered messages sent through Twilio in each experiment. }
\label{tab:error_table}
\end{table*}

\vspace{3mm}
\textbf{Complete disclosure:} To acquire bias-free results for the smishing study, we were deceptive to our participants. As a result, we had incomplete disclosure of our study. This may harm our participants by wasting their time or leading them to believe that they fell victim to a smishing attack. Since incomplete disclosure is being utilized, we informed the subjects of the true purpose of the study at the end via a debriefing message. Our debriefing message also educates our participants about actual smishing attacks. The debriefing message is provided in the Appendix. To minimize users' disruption, we sent a maximum of two messages to each user.

\vspace{3mm}
\textbf{IT department support:} Despite including a debriefing message for smishing victims at the end of our experiment, any participant may contact our university’s IT department. Before the experiment, we had a meeting with our university's IT department to discuss our study's potential security concerns. We got a letter of support from the IT department regarding the experiment. The support letter stated that if any participant contacted them regarding the possibility of being a victim, the IT department would assure them that we are conducting this study with all ethical considerations.

\vspace{3mm}
\textbf{Other ethical considerations:} We collected participants’ mobile numbers to send smishing messages and email addresses for a raffle draw with a chance to win Amazon gift cards. We also recorded participants’ text message responses and surveyed responses as part of our study. Likewise, we followed standard IRB protocols to mitigate the risks in this scenario. The participants’ responses and data are stored in a secure folder that can only be accessed by the authors.

\subsection{Recruitment and Demographics}
The participants in our study were primarily from the US and recruited through online channels. We posted recruitment advertisements to our social media accounts (e.g., LinkedIn and Facebook). We also requested different departmental chairs from our university to circulate our advertisements within their respective departments. To comply with legal requirements of the IRB and our own ethical considerations, we conducted our experiments in a manner where we did not collect any other personal information other than what was required (i.e., mobile number and email). We obtained consent from the participants before the experiments were conducted. After the consent, we collected the participants’ mobile phone numbers. We received a total of 1002 responses from the Qualtrics. After cleaning up our participant registrations by deleting invalid or duplicate participant registration entries, we had 622 available participant targets for testing our messages. During the first attempt, we sent out the SMSes to all 622 participant targets. We saw that around 56.27\% of messages were undelivered. 
In Table~\ref{tab:error_table} we show the error messages received for each experiment for the undelivered messages through Twilio. Error Code:30007 corresponds to messages being filtered by either Twilio or the carrier. Error code 30003, 30005 and 30006 were undelivered due to unknown or of unreachable destinations. For the second round of experiments, we specifically chose the targets which were successfully delivered in round 1 and included a random sample from the previously undelivered set to re-attempt. Over both rounds we were able to deliver at least one message to 265 unique participants. We note the difference in reasons for non-deliverability in round 2 compared to round 1, as the primary error code for round 2 was due to messages filtering by either the carrier or Twilio. We also collected demographic information such as age, gender, education level, profession, race, and mobile usage per week. The details are listed in Table~\ref{tab:demo} in Appendix. Our participants were comprised of 53.58\% male and 46.42\% female. Additionally, 54.71\% of participants were between the range of 25-34 years and have either an associate degree (25.28\%) or a bachelor’s degree (32.45\%). Furthermore, 34.72\% of participants use their phone between 6-10 hours of time per week. Our participants also have diverse professions and races. These participants were asked to provide an email to enter a raffle draw for a \$50 Amazon gift card for every 50 participants.

\subsection{Smishing Experiment} 
We designed and developed an automated control system to send multiple messages to our recipients at once, with only the click of a button. We used the bulk messaging service, Twilio, which provides an API endpoint. Furthermore, we used C\#, HTML, and JavaScript programming language to build the GUI part of our system. We conducted our experiments from November 25, 2021, to December 23, 2021. We sent messages during the weekdays and regular office hours. In the first round, we randomly selected a bunch of numbers to send one of the seven SMS. However, the user’s service provider (i.e., carrier) or Twilio filtering system blocked some messages and, thus, were undelivered. In the second round, we selected numbers from the first round, which showed a successfully delivered status in the Twilio ecosystem, along with a random selection of undelivered participants. A second round was necessary to measure the smishing rate of multiple attempts within a short span of time (e.g., a week). We ensured that we sent a different message to our recipients than the one that was previously sent.

\section{Performance Measures}
In this section, we analyze the CLICK, CALL, and REPLY rate of the data collected during the experiment.

\vspace{3mm}
\noindent \textbf{Performance metrics:}  In a smishing scenario, any response benefits the attackers, so we consider a participant responding to our message as a victim. Even in the scenario where a user might suspect that a message is a smishing attempt, responding will notify the attacker that the number is active, and that you are reading and considering the messages. This will invite more phishing messages in the futures. Additionally, by responding to the message a user may unknowingly provide personal information and increase their vulnerability to the social engineering attack~\cite{toClickOrNot,cybersecurityBehaviorIndia}. We define the \textit{Smishing Rate} as the proportion of users that click through an attacker’s page, reply through the text, or call a provided number. In the following section, we report experimental results for the three \textbf{RQs}.

\subsection{Potential Smishing Rate} To recall, our first research question (\textbf{RQ1}) was to find the potential success rate of smishing. We summarized our potential smishing success rate across 12 different experimental results in Table~\ref{table:summarresults}. From the table, we observe that we had potentially successfully convinced 16.92\% (44/260) of our participants in the first attempt to respond to our smishing campaigns. We found that the Facebook login scam message (E4) had the highest success rate with 34.62\% (9/26). Additionally, we observed that participants were also potentially falling for smishing during the second round. In the second attempt, we got an overall response rate of 12.82\% (25/195). Within this attempt, the message of an iPhone award from FedEx had the highest potential smishing success rate at 18.33\% (11/60). We did not observe any statistically significant difference between the smishing success rate in the first and second attempts. This may suggest that users are vulnerable to smishing irrespective of how many smishing messages they might have seen before.

The overall potential smishing success rate for messages with the reward (E1+E7+E8+E11+E12) scenario is 17.87\% (37/207) and the overall success rate for the messages with fear (E2+E3+E4+E5+E6+\newline E9+E10) scenario is 12.96\% (32/247). The potential smishing success rate for the messages with CLICK (E1+E4+E6+E8+E12) action is 18.18\% (34/187), the Reply (E2+E4+E7+E11) action is 14.47\% (22/152), Call(E3+E5+E9+E10) action is 9.22\% (13/141). 

\begin{table}[]
\centering
\small
\begin{tabular}{@{}lllll@{}}
\toprule
\textbf{Attributes(Baseline)} & \multicolumn{1}{c}{\textbf{\begin{math}\beta\end{math}}} & \multicolumn{1}{c}{\textbf{\begin{math}CI_{95\%}\end{math}}} & \textbf{T-value} & \textbf{\textit{p}-value} \\ \hline
{\ul{\textit{Area Code(vs local):}}} &  &  &  &  \\
Toll-free & -1.112 & [-2.67, 0.45] & -1.397 & .162 \\
Westchester, NY & -1.245 & [-3.24, 0.75] & -1.221 & .222 \\ \midrule
{\ul{\textit{User Action(vs CLICK):}}} &  &  &  &  \\
REPLY & .996 & [-0.95, 2.94] & 1.754 & .937 \\
CALL & -.035 & [-1.31, 1.24] & -0.070 & .348 \\
REPLY+CLICK & 2.348 & [0.30, 4.40] & 2.247 & .025* \\ \midrule
{\ul{\textit{Motivation(vs Fear):}}} &  &  &  &  \\
Reward & .378 & [-0.86, 1.62] & 0.598 & .550 \\ \midrule
{\ul{\textit{Entity(vs IRS):}}} &  &  &  &  \\
Walmart & 1.615 & [0.07, 3.16] & 2.049 & .041* \\
Bank & 1.160 & [-0.92, 3.24]  & 1.091 & .275 \\
Law Enforcement & -.433 & [-1.79, 0.93] & -0.624 & .532 \\
Facebook & 2.383 & [0.74, 4.03] & 2.844 & .004* \\
FedEx & .413 & [-0.56, 1.39] & .833 & .405 \\
Video website & .035 & [-1.24,1.31] & .054 & .957 \\
FRTUNE 500 Co & 1.409 & [-0.31,3.13] & 1.603 & .109 \\ \bottomrule
\end{tabular}
\caption{Omnibus Logistic regression of message attributes against the majority case Baseline. Scenario removed as it overlaps with the Entity. \begin{math}Adj. R^2 = 0.057.\end{math} * are statistically significant.}
\label{tab:AttributeComparison}
\end{table}

\vspace{3mm}

Our second research question (\textbf{RQ2}) was to understand the effect of different attributes on the success rate of smishing.  We conducted a Mann-Whitney U test, and Omnibus Logistic Regression to analyze the impact of different attributes on the smishing rate. To conduct this test, we first grouped the messages by scenarios -- fear and reward, and then separated into other attributes. We summarize the results of the Mann-Whitney U Test in Table~\ref{tab:hypothesis} and Logistic Regression results in Table~\ref{tab:AttributeComparison}.

\vspace{3mm}
\noindent
\textbf{Does the motivation have a significant impact on the smishing success rate?} The motivation behind a smishing message is how an attacker urges their victim to adopt some call to action. A smishing scenario primarily involves clicking on a link, replying to the message, or calling a number. We divided our smishing attempts into reward (37/207) and fear (32/247) motivated attacks and measured their overall success rates. We found a (\textit{p} = 0.550, \begin{math}CI_{95\%}\end{math} [-0.86, 1.62]), indicating that we cannot conclude that fear and reward-motivated smishing messages have statistically different impacts on smishing success rates.

\vspace{3mm}
\noindent
\textbf{Does the entity scenario have a significant impact on the smishing success rate?} Observations of real-world smishing reveal that there are a wide variety of fraudulent scenarios available for a scammer to leverage against victims. Often, well-known organizations or institutions are spoofed to trick a victim into trusting an attacker. We explore the differences that various scenarios have influencing smishing success. Our analysis of the success of these smishing attacks is based on the number of possibly tricked targets through their interactions with the messages. We conducted a right tailed p-approach between groups to measure the effect on smishing success rates.

In our reward motivating test, we compare the scam scenario of a gift card offer against a job offer at a Fortune 500 company. This paired test between groups E11 and E8 resulted in a [U = 937.500, \textit{p}  = .714]. Using our confidence level of 95\%, we did not find any statistical significance in the impact of an entity on the success rate of reward-based smishing attacks. Independently, we performed another entity scenario type test using fear-based motivation between groups E4 and E5. In this case, the scenario explored was a login event against suspicious financial activity made in the target's name. With the same right tailed p-approach, we found a [U = 648.00, \textit{p}  = .016], showing statistical significance of the entity on the potential success rate of fear-based smishing attacks using our 95\% confidence level. Additionally, when we compared the IRS entity scenario against other entities, we found that participants with Facebook(p = 0.004, \begin{math}CI_{95\%}\end{math} [0.74, 4.03]), and Walmart(p = 0.041, \begin{math}CI_{95\%}\end{math} [0.07, 3.16]) entities had a statistically significant impact on the odds of falling into smishing.

\vspace{3mm}
\noindent
\textbf{Does the area code have a significant impact on the smishing success rate?} To answer our research question, we first defined that our mobile phone area codes can be divided into two categories: local and toll-free numbers. A mobile-based scammer can choose either option to deliver their smishing message to their target. Toll-free numbers can be purchased for marketing purposes, while local numbers can be spoofed so that their area code appears as if they are local. To determine whether the area code significantly impacts smishing success rates, we used a confidence level of 95\%. Like previous tests, we split our experiments across reward and fear motivations to analyze the effect of both factors.

Using our reward-based motivation results, we performed a Mann-Whitney U-Test to compare local and toll-free to determine whether there were statistically significant differences. The results [U = 1154.500, \textit{p} = .409], indicate that we cannot conclude that the smishing messages that use a local number perform significantly different than toll-free ones. In our fear-motivated study, on the effect of area codes on smishing success rates, we found no significant difference with an [U = 549.000, \textit{p}  = .505].

\begin{table*}[]
\centering
\footnotesize
\begin{tabular}{@{}lcccccccl@{}}
\toprule
\textbf{Motivation} & \textbf{Type} & \textbf{GroupA} & \textbf{\#p-smishing} & \textbf{GroupB} & \textbf{\#smishing} & \textbf{U-Test} & \textbf{z} & \textbf{\textit{p}-value} \\ \midrule
\multirow{4}{*}{Reward} & Entity Scenario ( Fortune 500 Comapny vs Walmart) & E11 & 7/48 & E8 & 5/41 & 937.500 & -.366 & .714 \\
 & Area Code (Local vs Toll Free) & E12 & 11/60 & E8 & 5/41 & 1154.500 & -.826 & .409 \\
 & User Action ( CLICK vs REPLY) & E12 & 11/60 & E11 & 7/48 & 1361.500 & -.470 & .638 \\
 & Repeated smishing (Walmart) & E1 & 9/30 & E8 & 5/41 & 505.500 & -1.849 & .064 \\ \midrule
\multirow{4}{*}{Fear} & Entity (FB vs IRS) & E4 & 9/26 & E5 & 8/64 & 648.00 & -2.416 & .016* \\
 & Area Code (Toll Free vs Local) & E3 & 3/31 & E10 & 2/37 & 549.00 & -.667 & .505 \\
 & User Action ( CLICK vs CALL) & E6 & 4/30 & E5 & 8/64 & 952.00 & -.112 & .911 \\
 & User Action (CALL vs REPLY) & E5 & 8/64 & E2 & 6/50 & 1592.00 & -.080 & .936 \\ \bottomrule
\end{tabular}
\caption{Summary of the Mann-Whitney U Test results.}
\label{tab:hypothesis}
\end{table*}

\vspace{3mm}
\noindent
\textbf{Does the response method have a significant impact on the smishing success rate?}
Another attribute that we explored in this study is the effects of user prompted actions on the success rate of smishing attacks. In this study, we separate our action types into the CLICK, CALL and REPLY. While these three user action types can be further divided into more specific types, we compared their broader forms through individual and group-based right-tailed significance tests. 

In our first statistical test, we compared the smishing messages which incorporate a CLICK vs REPLY user action request. In these smishing attacks, we define a successful click-based smishing attack as one in which the target clicks on the link in the message, while a successful REPLY attack requires the user to respond to the message [U = 1361.500, \textit{p}  = .638], showing no significant difference between the two user actions.

For fear-motivated smishing messages, we have two sets of experiments to compare, CLICK vs CALL and CALL vs REPLY. For our CLICK vs CALL test, we identified a successful call-based smishing attack as the victim calling the number provided in the smishing message. Likewise, CLICK vs REPLY did not yield any statistically significant difference between the user action in the smishing experiment with an [U = 952.000, \textit{p}  = .911].

Our final individual user action test compared CALL vs REPLY with a fear-based motivated smishing message. Using the same right-tailed significance test we again did not find any statistically significant difference between CALL and REPLY user actions with an [U = 1592.000, \textit{p}  = .936]. Based on these tests we cannot conclude that CLICK, CALL or REPLY have a statistically significant impact on the smishing success rate. 

While our individual statistical tests of user action did not yield any significance between experiment groups, we also compared overall CLICK+REPLY, CALL, and REPLY actions to a CLICK user action baseline. We found that odds of participants who received CLICK+REPLY messages had a statistically significant(p = 0.025, \begin{math}CI_{95\%}\end{math} [0.30, 4.40]) increase in the odds of potentially falling for our smishing messages than those with just CLICK.

\vspace{3mm}
\noindent
\textbf{Does repeated smishing have a significant impact on the smishing success rate?} A research question that we considered concerned whether victims of past smishing attacks were more likely to fall into future attacks. The hope is that when victims have fallen into smishing, they learn from the experience and are careful about future smishing attacks. To measure the effects of repeated smishing, we used the same reward-based Walmart smishing message for both of our experiments. With a [U = 505.500, \textit{p}  = .064], we did not find that repeated smishing has a statistically significant effect on smishing rates.

\begin{table}[ht]
\centering
\small
\begin{tabular}{@{}cccc@{}}
\toprule
\textbf{Experiment} & \textbf{Stop option} & \textbf{Stop Count} & \textbf{\%} \\ \midrule
E7 & Yes & 4 & 18.18 \\ \midrule
E6 & No & 4 & \multirow{3}{*}{31.82} \\
E5 & No & 2 &  \\
E3 & No & 1 &  \\ \bottomrule
\end{tabular}
\caption{The participant proactively replied "STOP" while we did not have that option in the messages.}
\label{tab:stopcount}
\end{table}

\vspace{3mm}
\noindent
\textbf{The importance of security awareness} 
For our study, we assumed an attacker was launching a smishing attack to find active phone numbers, share smishing links, gather private information, and accordingly crafted messages. When falling victim to mass smishing, opening and replying to a malicious text message can alert an attacker that the targeted phone number is still active and to initiate further targeting. In response to smishing, participants tended to adopt the stereotypical belief that replying STOP will stop the hackers. Based on Table~\ref{tab:stopcount}, we have seen that many participants thought that a messaging STOP will stop the smishing. However, this is a bait for the attackers to find active numbers. During our study, we received 11 STOP replies out of 22 text replies. The participants even proactively replied STOP 31.82\% of the time, even when no STOP option was mentioned in the SMS. Our findings suggest that smishing training could support user knowledge and awareness regarding smishing attacks. To prevent smishing attacks, users need more cybersecurity education, and the research community should develop better techniques to protect against those smishing attacks.

\vspace{3mm}
\noindent
\subsection{Impact of Demographics on the smishing success rate}
Our third research question (\textbf{RQ3}) was to find the effect of demographic factors on the success rate of smishing. Different studies have found different human attributes that are more prone to smishing attacks. Sheng et al.~\cite{sheng2010falls} found that women and participants between 18 to 25 are more susceptible to smishing attacks than other participants. Similarly, authors in~\cite{williams2018exploring} found that the age range between 18 and 25 are more susceptible to phishing than any other age. Lin et al.~\cite{lin2019susceptibility} revealed that older women are the most vulnerable group. Hadlington et al.~\cite{hadlington2017human} found that men are more susceptible to mobile smishing attacks than women.

We summarized the demographic statistics of our study in Table~\ref{table:demoresults} and the results of our Logistic regression in Table~\ref{tab:demographicsLogistic}. Due to the nature of CLICK replies not being tied to individual participants, we measure only CALL and REPLY user actions again our demographics data. In terms of age group, individuals aged 45-54 have a higher percentage of potentially falling into smishing scams than any other age group. The 18-24 age group has the lowest smishing success rate at 4.17\%. It is assumed that doctoral-level students should have higher domain knowledge than high school students. Based on our study, we found that participants with a doctoral degree have a 37.50\% potential smishing success rate, which is nearly double that of participants with less than a high school degree.

\begin{table}[]
\footnotesize
\centering
\begin{tabular}{@{}cccl@{}}
\toprule
\textbf{} & \textbf{Range} & (\textbf{\#Smish/N}) & \multicolumn{1}{r}{\textbf{(\%)}} \\ \midrule
\multirow{2}{*}{Gender} & Male & (14/142) & 9.86 \\
 & Female & (14/123) & 11.38 \\ \midrule
\multirow{4}{*}{Age (in years)} & 18-24 & (2/48) & 4.17 \\
 & 25-34 & (17/145) & 11.72 \\
 & 35-44 & (6/60) & 10.00 \\
 & 45+ & (3/12) & 25.00 \\ \midrule
\multirow{7}{*}{Education} & \textless High School & (1/5) & 20.00 \\
 & High School degree & (3/26) & 11.54 \\
 & Associate degree & (7/67) & 8.96 \\
 & Some college & (6/49) & 12.24 \\
 & Bachelor's degree & (7/86) & 8.14 \\
 & Master's degree & (1/24) & 4.17 \\
 & Doctoral degree & (3/8) & 37.50 \\ \midrule
\multirow{8}{*}{Ethnicity} & White & (10/128) & 7.81 \\
 & African American & (9/46) & 19.57 \\
 & Asian American & (2/28) & 7.14 \\
 & Mexican American & (5/23) & 21.74 \\
 & American Indian & (0/22) & 0 \\
 & Other His./Lat. & (0/13) & 0 \\
 & Native Hawaiian & (2/3) & 66.67 \\
 & Other & (0/2) & 0 \\ \midrule

 \multirow{5}{*}{\begin{tabular}[c]{@{}c@{}}Mobile Usage \\ (in hours/week)\end{tabular}} & 1-5 & (1/23) & 4.35 \\
 & 6-10 & (6/92) & 6.52 \\
 & 11-20 & (12/77) & 15.58 \\
 & 21-30 & (7/46) & 15.22 \\
 & \textgreater31 hours & (2/27) & 7.41 \\ \midrule
\multirow{12}{*}{Occupation} & Student & (6/52) & 11.54 \\
 & Management professional & (5/40) & 12.50 \\
 & Sales and Office & (5/40) & 12.50 \\
 & Technician & (0/31) & 0 \\
 & Service & (4/23) & 17.39 \\
 & Software Engineer & (2/22) & 9.09 \\
 & Teacher & (1/14) & 7.14 \\
 & Construction and Maintenance & (1/14) & 7.14 \\
 & Farming fishing and forestry & (2/8) & 25.00 \\
 & Production transportation & (1/8) & 12.50 \\
 & Government & (1/4) & 25.00 \\
 & Other & (0/9) & 0 \\ \bottomrule
\end{tabular}
\caption{Survey participant demographics and their respective smishing success percentage.}
\label{table:demoresults}
\end{table}

\begin{table}[]
\centering
\scriptsize
\begin{tabular}{@{}ccclc@{}}
\toprule
\textbf{Demographic(Baseline)} & \multicolumn{1}{c}{\textbf{\begin{math}\beta\end{math}}} & \multicolumn{1}{c}{\textbf{\begin{math}CI_{95\%}\end{math}}} & \textbf{T-value} & \textbf{\textit{p}-value} \\ \midrule
\multicolumn{1}{l}{{\ul{ \textit{Gender (vs Male):}}}} & \multicolumn{1}{l}{} & \multicolumn{1}{l}{} &  & \multicolumn{1}{l}{} \\
Female & 0.090 & [-0.98, 1.16] & \multicolumn{1}{c}{0.165} & 0.870 \\ \midrule
\multicolumn{1}{l}{{\ul{\textit{Age(vs 25-34):}}}} &  &  &  &  \\
18-24 & -3.338 & [-5.67,-1.00] & -2.800 & 0.005* \\
35-44 & -.392 & [-1.89, 1.11] & -0.512 & 0.608 \\
45+ & .573 & [-1.62, 2.76] & 0.513 & 0.608 \\ \midrule
\multicolumn{1}{l}{{\ul{\textit{Education(vs Bachelors degree):}}}} & \multicolumn{1}{l}{} & \multicolumn{1}{l}{} &  & \multicolumn{1}{l}{} \\
\textless High School & .137 & [-2.84, 3.11] & 0.090 & 0.928 \\
High school graduate & .466 & [-1.41, 2.35] & 0.486 & 0.627 \\
Some college & .112 & [-1.31, 1.53] & 0.155 & 0.877 \\
Associate degree & -.429 & [-1.83, 0.98] & -0.598 & 0.549 \\
Masters degree & 1.510 & [-1.21, 4.23] & 1.086 & 0.277 \\
Doctoral degree & -.999 & [-3.73, 1.73] & -0.717 & 0.473 \\ \midrule
\multicolumn{1}{l}{{\ul{\textit{Ethnicity(vs White):}}}} &  &  &  &  \\
African American & 1.879 & [0.51, 3.25] & 2.696 & 0.007* \\
Asian American & .374 & [-1.93, 2.67] & 0.319 & 0.750 \\
Mexican American & 2.368 & [0.42, 4.31] & 2.387 & 0.017* \\
American Indian & -18.132 & [-14901, 14864] & -0.002 & 0.998 \\
Other His./Lat. & -19.000 & [-19343, 19305] & -0.002 & 0.998 \\
Native Hawaiian & 3.604 & [-0.93, 8.14] & 1.558 & 0.119 \\
Other & -18.463 & [-51004, 50967] & -0.001 & 0.999 \\ \midrule
\multicolumn{1}{l}{{\ul{\textit{Mobile Usage(vs 6-10):}}}} & \multicolumn{1}{l}{} & \multicolumn{1}{l}{} &  & \multicolumn{1}{l}{} \\
1-5 hours & -.346 & [-3.04, 2.35] & -0.252 & 0.801 \\
11-20 hours & 2.208 & [0.67, 3.75] & 2.801 & 0.005* \\
21-31 hours & 1.493 & [-0.10, 3.09] & 1.834 & 0.670 \\
31+ hours & 1.048 & [-1.05, 3.14] & 0.981 & 0.327 \\ \midrule
\multicolumn{1}{l}{{\ul{\textit{Profession(vs Student):}}}} &  &  &  &  \\
Management professional & -1.542 & [-3.55, 0.47] & -1.506 & 0.132 \\
Sales and office & -1.183 & [-3.19, 0.82] & -0.983 & 0.247 \\
Technician & -20.322 & [-12467, 12427] & -0.003 & 0.997 \\
Service & -.366 & [-2.52, 1.79] & -0.332 & 0.739 \\
Software Engineer & -2.527 & [-5.62, 0.57] & -1.601 & 0.109 \\
Teacher & -20.401 & [-21959, 21919] & -0.002 & 0.999 \\ 
Construction and maintenance & -1.761 & [-4.54, 1.02] & -1.242 & 0.214 \\
Farming fishing and forestry & -.556 & [-3.29, 2.18] & -0.398 & 0.691 \\
Production transportation & -.826 & [-3.87, 2.22] & -0.532 & 0.595 \\
Government & -1.124 & [-4.37, 2.12] & -0.679 & 0.497 \\
Other & -2.478 & [-5.47, 0.52] & -1.601 & 0.105 \\ \bottomrule
\end{tabular}
\caption{Omnibus Logistic Regression of demographics with respect to potential smishing rates. Variables are categorically compared using the majority case as the Baseline.  \begin{math}Adj. R^2 = 0.404.\end{math} * are statistically significant.}
\label{tab:demographicsLogistic}
\end{table}

In terms of profession, government employees and participants in the farming, fishing, and forestry had the highest potential smishing success rate with 25\% followed by the service holders with 17.39\%. It is worth mentioning that technicians comprised of 11.7\% of the participants and no one had potentially fallen victim to any of our smishing attempts. Concerning race, Native Hawaiians had the highest smishing success rate with 66.67\%, although they comprised of only 1.1\% of our participants’ pool. Mexican Americans and African Americans had a smishing success rate of 21.74\% and 19.57\% respectively. People who spent 11-20 hours on the mobile had the highest smishing success rate at 15.58\%, followed by people who spent 21-30 hours per week.

\textbf{Statistical analysis on demographics} This analysis was used to study the relationship between our smishing results and the demographic parameters of our participants. The statistic analysis is done on a unique list of participants who had successfully been delivered a message in one of the study's experiments. We provide the statistical analysis with different demographics parameter as our independent variables against our potential smishing success rate dependent variable. To test our categorical demographic data, we use the majority case of each variable for the reference. 

\vspace{3mm}
\noindent
\textbf{Age} We found that 18-24 year old participants had statistically significant(\textit{p} = 0.005, \begin{math}CI_{95\%}\end{math} [-5.67, -1.00]) lower odds in potentially falling for our smishing messages. 

\vspace{3mm}
\noindent
\textbf{Gender} We did not find any statistical significant difference in the odds of potentially fall into a smishing scam when comparing Female to Male participants.

\vspace{3mm}
\noindent
\textbf{Education Level} There was no statistical significant difference observed in the odds of participants potentially falling for our smishing smishing scam based on different education levels.

\vspace{3mm}
\noindent
\textbf{Race} We found that in comparison to our majority case, African American participants had a statistically significant increase (\textit{p} = 0.007, \begin{math}CI_{95\%}\end{math} [0.51, 3.25]) in the odds of potentially falling victim our smishing attempt. Participants that identified as Mexican Americans also had a statistically significant(\textit{p} = 0.017, \begin{math}CI_{95\%}\end{math} [0.42, 4.31]) increase in the odds of potentially falling victim to our smishing scams. 

\vspace{3mm}
\noindent
\textbf{Profession} We see no statistically significant difference in the potential smishing rates for participants of different professions when compared to our Student baseline. 

\vspace{3mm}
\noindent
\textbf {Mobile usage} Based on our results, we found that using the phone 11-20 hours per week had a statistically significant increase(\textit{p} = 0.005, \begin{math}CI_{95\%}\end{math} [0.67, 3.75]) in the odds of potentially falling for a smishing scam, compared to our 6-10 hour baseline.

\section{Survey Responses}
\label{sec:survey}
In this section, we provide the recorded responses for the post-experiment survey. In comparison with the study by Tu et al.~\cite{tu2019users}, we found a higher survey response rate of 1.76\%(8/454) against their rate of 1.17\%(35/3000). We received 2 survey responses in Qualtrics and 12 voice calls, but 9 were either silent or contained no useful information. Therefore, we received 3 meaningful call feedback responses. Concerning text replies, we received 21 but a few of them were blank or contained memes. In 5 experiments we received meaningful survey responses, which we summarized in Table~\ref{table:reasonsforresponse}. For all click experiments, we got a total of 44 clicks; however, we only received 1 response for E1 and 1 response for E12. Despite participants falling for smishing attacks 69 times during this study, post-experiment feedback rates were low.

Most of the survey respondents were aware of the scam, yet they responded. In the smishing context, any response is valuable to hackers. A person who was convinced by smishing mentioned that due to the local number, important government information, and a tax lawsuit, they fell into the smishing scam.

Regarding our IRS fear-motivated message, we received several meaningful call responses. These responses indicated that they fell for the message because they felt it was realistic. Factors included that the message was coming from a local number, that the message was threatening and that they responded because the topic related to important government information. However, despite the larger number of survey responses, we found that our Facebook fear-motivated messagem (M4 in Table~\ref{table:messagingcontent}) had a significantly higher smishing success rate. This indicates that those who believed the IRS-based smishing message was authentic were more likely to provide a post-survey response.

While we only received two survey responses to our click-based experiments, the targets of our smishing campaign indicated curiosity as their main factor for potentially falling for the smishing scam. As per PhishMe~\cite{phishme}  2017, curiosity is one of the top motivators of smishing attacks. Whether different user-action smishing attacks are motivated by different factors is work we leave for future studies. One victim mentioned that they recognized the message as smishing and clicked anyway to see what would happen. This action is inadvisable even if the target knows that they must be careful. They should not be visiting malicious websites. One user mentioned that they were able to preview the message to see where it would go before clicking. This is a useful feature that is available on modern mobile devices. However, smishing attacks often spoof well-known websites, thereby making them indistinguishable from the authentic one.

\begin{table*}[]
\centering
\small
\begin{tabular}{|l|l|l|l|}
\hline
\multicolumn{1}{|c|}{\textbf{No.}} & \textbf{Action} & \multicolumn{1}{c|}{\textbf{Reasons for response}} & \textbf{Aware?} \\ \hline
E1 & CLICK & Curiosity. & Yes \\ \hline
\multirow{2}{*}{E2} & \multirow{2}{*}{REPLY} & Because we do not spend that kind of money at any Walmart & Yes \\ \cline{3-4} 
 &  & Credit card fraud & Yes \\ \hline
\multirow{3}{*}{E5} & \multirow{3}{*}{CALL} & \begin{tabular}[c]{@{}l@{}}The fact that the area code for the phone number was the city of my school. So it made it \\  like more realistic, rather than a random area code from a different city-state or country.\end{tabular} & No \\ \cline{3-4} 
 &  & \begin{tabular}[c]{@{}l@{}}Important government information, just because it has to do with taxes, it has to do with \\ security numbers and all kinds of personal information.\end{tabular} & No \\ \cline{3-4} 
 &  & The fact that you were threatening me about my tax information. & No \\ \hline
E7 & REPLY & If this is legit where do you get your contact list from? & Yes \\ \hline
E12 & CLICK & \begin{tabular}[c]{@{}l@{}}I often find them funny, so I still open them even if I know it's fake. Also helps me\\  educate myself on how these cons work.\end{tabular} & Yes \\ \hline
\end{tabular}
\caption{Summary of valid survey responses.}
\label{table:reasonsforresponse}
\end{table*}

\section{Discussion}
After analyzing the results of this study, we are convinced that nicely crafted smishing messages can be very effective for smishing attacks. In experiment E4, with the fear of losing social media accounts, we received the highest smishing success rate of 34.62\% (9/26). In this case, we used M4 which prompts urgency in the message. The second and third highest smishing success rates are 30.00\% (9/30) for experiment E1 and 18.33\% (11/60) for E12, correspondingly. In the message, we mentioned the chance to win a \$500 Walmart gift card in E1 and an iPhone winning message in E11. We obtained a 12.82\% smishing success rate in the second attempt compared to 16.92\% in the first. The participants proactively responded 7 times in the first round (* in Table~\ref{table:summarresults}) compared to 2 times in the second round. These findings imply that participants were aware of the smishing attacks in the second attempt.

\textbf{Similarities and differences of Robocalls study} Tu et al.~\cite{tu2019users} conducted a detailed study regarding Robocall scams. This study inspired our work. There are several similarities with Robocalls study, such as examining the effects that the area codes, entity, and motivation have on the success rate of smishing attacks. In terms of motivations, we compared our individual feature types against both fear and reward messages. Regarding entity scenarios, the difference between our study and theirs is that we changed the entity scenario within the text, and they changed the entity within the caller’s name, mimicking caller ID spoofing. Our results were only statistically significant in our fear-motivated results when comparing entity scenarios, while theirs were educationally significant overall.

Concerning the area code type smishing messages, Tu et al. found that toll-free area codes had a small impact on the success rate of phishing attempts. However, in our study, we did not find any statistical difference when comparing the area code for either reward-motivated or fear-motivated attacks. While comparing the effects that motivation had on our smishing success rate, we found similar results to Tu et al. In an overall comparison of all our fear-based and reward-based messages, we did not find any statistically significant difference between the two motivation options. Subsequently, we were not able to conclude whether the choice between fear or reward motivated attacks had an impact on phishing rates.

\textbf{Potential defense mechachism} In our study, we observed that Twilio or Telecommunication Carriers potentially filtered some of our SMS. Therefore, further research is required to understand the blocking scheme of current SMS systems. Our study findings suggest that the current smishing blocking system is not protecting the users enough against smishing. The automatic smishing detection system can also be deployed in the Bulk SMS senders and carriers. Additionally, there could be a third-party block system.

\textbf{Study limitations} In our study, we measure the efficacy of smishing in a realistic scenario. However, there are several limitations to our study. The first limitation is that we must obtain consent from the user before conducting the study for ethical reasons. Participants may recall consenting to a behavioral study for mobile users and giving out their phone numbers a week before the experiment. We believe that this may bias the results. Our results could have been more impartial had we not received consent from participants like in the previous telephone scam study~\cite{tu2019users}. However, for us to conduct these experiments, we had to collect mobile phone numbers before the experiment and obtain proper consent from users to comply with the IRB guidelines. 

Secondly, we considered the number of delivered and undelivered messages based on the status of the bulk messaging service (i.e., Twilio). Through Twilio, we can discern the status of our messages as sent, delivered and undelivered. In this study, a delivered status is interpreted as a receipt from the carrier regarding the confirmation of message delivery. However, there are conditions where a message may still not arrive in the participants inbox. Third-party anti-smishing apps and device or carrier issues may result in a failed delivery. For the purposes of this study we consider only delivered messages; however, this could potentially include false positives.

Thirdly, the number of populations in our experiment might be low. However, we have a wide variety of population in our study. While the participants’ demographic is diverse, they are not census representative of the adult 18+ population. Next, there might be confounding factors by which our messages were delivered that could impact bias. Messages were sent during regular office hours throughout November 25, 2021, to December 23, 2021. We acknowledge that the time of the year may have affected response rates. Lastly, as the content of messages must change between different entities, scenarios, user actions and motivations, they do not lend themselves to perfect comparisons. In other words, as these factors change, so do the composition of our messages. The changes in the composition of the messages may lead to additional biases. To limit this bias, we intend to select test groups for comparisons based on one attribute difference.

\section{Related Work} Based on the existing literature, very limited empirical study has been done on mobile fraud. We did not find any systematic and thorough empirical study for smishing. The most relevant smishing study was conducted by Blancaflor et al.~\cite{blancaflor2021let}. They conducted a phishing campaign with one phishing link to learn the types of phishing attacks users are most susceptible to. They sent that phishing link to 107 people - 46 via email, 37 via social media chat, and 24 via SMS. They have found that only 1 out 24 people click the phishing link.  Moreover, they created a framework to measure user interaction with phishing scams. Their results demonstrated that trust was the most important factor for why users click on phishing links. The other relevant study on mobile fraud was an empirical study published by Tu et al.~\cite{tu2019users}. However, this study mainly focused on Robocalls. More specifically, they pre-recorded their messages and automated calls to specific numbers to ask for the last 4-digit of their social security number. Their experiments also had multiple attributes like caller ID, voice production, gender, accent, and scenario.

Other relevant spam SMS studies mostly focus on detecting spam SMS using the classification method or uncover the fake base station for a spam campaign. Reaves et al.,~\cite{reaves2016detecting} developed a classification technique for detecting spam SMS with a precision of 100\% and recall of 96.8\%. Li et al.,~\cite{li2017fbs} developed a system that can detect fraud SMS messages by detecting Fake Base Station (FBS). Zhang et al., ~\cite{zhang2020lies} uncovered the characteristics of FBS spammers, including their business categories, temporal patterns, spatial patterns, and construction of fraudulent messages for a campaign. Chorge et al.~\cite{chorghesurveymobilephish} compared the detection rates of ML techniques to classify SMS phishing messages. Furthermore, Waters et al.~\cite{whytrustmessages} discussed physiological models that explain why users establish and maintain trust in messaging, and how this can explain human behavior relating to phishing scams.

Desolda et al.~\cite{desoldahumanfacotrsinphishing} explores the human factors in falling victim to phishing attacks. Their findings conclude that lack of knowledge and resources are the top two human vulnerabilities which lead to phishing. Additionally, Wash et al.~\cite{washphishingstories} analyzed the effectiveness of user training to address vulnerabilities to phishing and reduce users click-through rates on phishing links. Moul K. contributed by describing how users can spot social engineering attacks such as phishing~\cite{moulavoidphishingtraps}.

Similarly, many studies have been conducted on describing mobile fraud and the types of phishing attackers use. Razaq et al.~\cite{Razaqetallmobilefraud} described mobile financial fraud techniques used to gain access to a victim's pin or install malware through SMS phishing. Additionally, this work conducted interviews with victims of mobile fraud to discuss their understanding of mobile fraud and the motivations behind the attacker’s actions. Leonov et al.~\cite{leonovsocialengineeringtechniques} highlights smishing and the statistics on the types of phishing attacks. This work identifies scams involving spoofing online banking, stores and social media among the largest types of phishing attacks. Findings by Merwe et al.~\cite{merwephishinthesystem} on SMS phishing messages provides details on phishing scenarios and how an attacker can perform large-scale SMS phishing.

Many researchers have conducted email or website-based empirical phishing studies and found the relationship of phishing detection between different demographic factors. Several studies have mentioned age and gender as factors relating to phishing susceptibility~\cite{jagatic2007social,li2020experimental,oliveira2017dissecting,sheng2010falls}. Dhamija et al.,~\cite{dhamija2006phishing} conducted a usability study in which 22 participants were shown 20 websites and asked to determine which one was fraudulent. Their results indicated that users do not perform well at phishing detection and make incorrect choices 40\% of the time. Sheng et al.,~\cite{sheng2010falls} conducted an online survey of 1001 participants and found that women are more susceptible to phishing than men. They also found that ages 18 and 25 are more susceptible than other age groups. Due to the different methods and populations, Li et al.~\cite{li2020experimental} concluded that older people were the most susceptible. Several studies~\cite{butavicius2017understanding,flores2015investigating,harrison2016individual,sheng2010falls} have reported that phishing susceptibility is significantly related to user’s security knowledge, awareness, behavior, and previous anti-phishing training experience. In addition, several studies mentioned that employment department and position~\cite{li2020experimental}, and education level~\cite{moody2017phish,parsons2013phishing} were related to phishing susceptibility. The personality traits (i.e., The Big Five)~\cite{baki2017scaling,halevi2015spear} and cognitive impulsivity~\cite{butavicius2016breaching,parsons2013phishing} are also related to phishing susceptibility. An analysis of previous studies by Baki et al.~\cite{Bakisixteenyearsofstudies}, describes this inconsistency among conclusions on the demographic factors in phishing susceptibility. A common factor was the lack of comprehensive reporting. Their meta-analysis of over 80 research papers showed that males and older users performed better at identifying phishing emails.
\section{Conclusion} Smishing is a cybersecurity threat that is on the rise. In recent years, smishing has surpassed the number of spam calls. In our paper, we conducted a systematic and thorough empirical study on smishing. Accordingly, we presented methodology, design, procedure, and analyses to investigate how smishing works and provided real-world smishing results from smishing scams performed on 265 unique users. With the consent of our participants, we collected their numbers and conducted twelve different experiments. We found that our participants potentially fell into smishing across user action, Entity scenario, and several demographic features when compared to baseline values. Due to curiosity, more knowledgeable users may also become victims of smishing scams. To combat smishing, there is a need for greater user awareness and automatic SMS phish detection mechanisms.

\bibliographystyle{plain}
\bibliography{main}

\clearpage
\twocolumn
\section*{Appendix}
\appendix
\begin{table}[h]
\centering
\small
\begin{tabular}{@{}cccl@{}}
\toprule
\textbf{} & \textbf{Range} & \textbf{N} & \multicolumn{1}{r}{\textbf{(\%)}} \\ \midrule
\multirow{2}{*}{Gender} & Male & 142 & 53.58 \\
 & Female & 123 & 46.42 \\ \midrule
\multirow{4}{*}{Age (in years)} & 18-24 & 48 & 18.11 \\
 & 25-34 & 145 & 54.71 \\
 & 35-44 & 60 & 22.64 \\
 & 45 or older & 12 & 4.5 \\ \midrule
\multirow{7}{*}{Education} & \textless High School & 5 & 1.88 \\
 & High School degree & 26 & 9.81 \\
 & Associate degree & 67 & 25.28 \\
 & Some college & 49 & 18.49 \\
 & Bachelor's degree & 86 & 32.45 \\
 & Master's degree & 24 & 9.05 \\
 & Doctoral degree & 8 & 3.01 \\ \midrule
\multirow{8}{*}{Race} & White & 128 & 48.30 \\
 & African American & 46 & 17.35 \\
 & Asian American & 28 & 10.57 \\
 & Mexican American & 23 & 8.67 \\
 & American Indian & 22 & 8.30 \\
 & Other His./Lat. & 13 & 4.90 \\
 & Native Hawaiian & 3 & 1.13 \\
 & Other & 2 & 0.75 \\ \midrule
\multirow{5}{*}{\begin{tabular}[c]{@{}c@{}}Mobile \\ Usage\end{tabular}} & 1-5 & \multicolumn{1}{l}{23} & 8.68 \\
 & 6-10 & \multicolumn{1}{l}{92} & 34.72 \\
 & 11-20 & \multicolumn{1}{l}{77} & 29.06 \\
 & 21-30 & \multicolumn{1}{l}{46} & 17.36 \\
 & \textgreater{}31 & \multicolumn{1}{l}{27} & 10.19 \\ \midrule
\multirow{12}{*}{Occupation} & Student & 52 & 19.62 \\
 & Management professional & 40 & 15.09 \\
 & Sales and Office & 40 & 15.09 \\
 & Technician & 31 & 11.70 \\
 & Service & 23 & 8.68 \\
 & Software Engineer & 22 & 8.30 \\
 & Teacher & 14 & 5.28 \\
 & Construction and Maintenance & 14 & 5.28 \\
 & Farming fishing and forestry & 8 & 3.02 \\
 & Production transportation & 8 & 3.02\\
 & Government & 4 & 1.51 \\
 & Other & 9 & 3.40 \\ \bottomrule
\end{tabular}
\caption{Demographics of the survey sample (n=265).}
\label{tab:demo}
\end{table}

\textbf{Follow up debriefing Message}

Hello, I am an assistant professor [Anonymous]. We are conducting a research study to measure the efficacy of text
phishing although we advertised it as a behavioral study of mobile users. The action you performed could
potentially lead you to become exploited in a real text phishing scam. However, it was not an actual text
phishing scam and we did not collect any of your personal details in this experiment. We would appreciate
it if we get your response to our quick One-question survey along with your number. Thank You!

\textbf{Survey Question}

Survey Question for “TEXT": Could you please help us understand what was the most important factor that convinced you to reply to this message?

Survey Question for “WEBSITE”: Could you please help us understand what was the most important factor that convinced you to click on the link?

Survey Question for “CALL BACK”: Could you please help us understand what was the most important factor that convinced you to call back? When you are finished, please press the pound key to end recording.

\textbf{Ending Message}

Thank you. This is the end of the research experiment. If you have any questions concerning the research
study, please contact the research team at [Anonymous]. If you have any questions about your
rights as a participant in this research, or if you feel you have been placed at risk, you can contact the Chair
of the Human Subjects Institutional Review Board, through the university email at [Anonymous] or call (XXX) XXX-4029. Thank you for your participation. Goodbye!

\end{document}